\documentclass[aps,preprint]{revtex4}
\usepackage{epsfig}
\usepackage{wrapfig}
\begin{document}
\topmargin=0.2in
          
\title{A Measurement of the Average Longitudinal Development Profile of 
Cosmic Ray Air Showers 
Between ${\bf 10^{17}}$eV and ${\bf 10^{18}}$eV} 
\author{T.Abu-Zayyad$^1$, K.Belov$^1$, D.J.Bird$^{11}$, 
 J.Boyer$^{4}$, Z.Cao$^1$, M.Catanese$^3$, G.F.Chen$^1$,
 R.W.Clay$^{5}$, C.E.Covault$^2$,  H.Y.Dai$ ^1$,
 B.R.Dawson$^5$, J.W.Elbert$^1$, B.E.Fick$^2$, L.F.Fortson$^{2a}$,
 J.W.Fowler$^2$, K.G.Gibbs$^2$, M.A.K.Glasmacher$^7$,
 K.D.Green$^2$, Y.Ho$^{10}$, A.Huang$^1$ , C.C.Jui$^1$, M.J.Kidd$^6$,
 D.B.Kieda$^1$, B.C.Knapp$^4$, S.Ko$^1$, C.G.Larsen$^1$,
 W.Lee$^{10}$, E.C.Loh$^1$, E.J.Mannel$^4$, J.Matthews$^9$,
 J.N.Matthews$^1$ , B.J.Newport$^2$, D.F.Nitz$^8$,
  R.A.Ong$^2$, K.M.Simpson$^5$, J.D.Smith$^1$,
 D.Sinclair$^6$, P.Sokolsky$^1$, C.Song$^{10}$, 
 J.K.K.Tang$^1$, S.B.Thomas$^1$, J.C.van der Velde$^7$,
 L.R.Wiencke$^1$, C.R.Wilkinson$^5$, S.Yoshida$^1$ and
 X.Z.Zhang$^{10}$ }

\affiliation{$^1$ High Energy Astrophysics Institute, University of Utah, Salt Lake 
 City UT 8
 4112 USA\\$^2$ Enrico Fermi Institute, University of Chicago, Chicago IL 
 60637 USA\\
 $^3$ Smithsonian Astrophys. Obs., Cambridge MA 02138 USA\\
 $^4$ Nevis Laboratory, Columbia University, Irvington NY 10533 USA\\
 $^5$ University of Adelaide, Adelaide S.A. 5005 Australia \\
 $^6$ University of Illinois at Champaign-Urbana, Urbana IL 61801 USA\\
 $^7$ University of Michigan, Ann Arbor MI 48109 USA\\
 $^8$ Dept. of Physics, Michigan Technical University, 
 Houghton, MI 49931 USA\\
 $^9$ Dept. of Physics and Astronomy, Louisiana State 
 University, Baton Rouge LA 70803 and \\
 Dept. of Physics, Southern University, Baton Rouge LA 70801 USA\\
 $^{10}$ Dept. of Phys., Columbia University, New York NY 10027 USA\\
$^{11}$ Defence Science and Technology Organisation, P.O. Box 1500, Salisbury, S.A.  5108, AUSTRALIA\\
 $^a$ joint appt. with The Adler Planetarium and Astronomy Museum, 
 Astronomy Dept., Chicago IL 60605 USA\\ } 
\date{\today}

\begin{abstract}
The average extensive air shower longitudinal development profile 
as a function of shower age in the energy range from 
10$^{17}$ to 10$^{18}$eV is measured using data from 
 the hybrid HiRes/MIA experiment. An angular bin signal based Cerenkov light 
component subtraction method and a shower maximum fitting method
using a local parabolic function are used to correct and normalize the data.
The Gaisser-Hillas and Greisen functions work equally well for describing 
the shower profile. The Gaussian function is a poor fit. A simple 
scale-free function is proposed and fits the data equally well. The best-fit
parameters for the above functions are determined.
\end{abstract} 
\maketitle


\section{Introduction}

   The atmospheric nitrogen fluorescence light technique plays an increasingly
important role in extremely high energy cosmic ray observation. The success of
the Fly's Eye and HiRes\cite{hires} experiments encourages the employment of
this technique in future projects like the Auger experiment\cite{Auger}, 
the Telescope Array proposal\cite{TA} and space based experiments like EUSO
\cite{EUSO} and OWL\cite{OWL}.  In a recent
report on the change of cosmic ray composition in the vicinity of
5$\times$10$^{17}$ eV from HiRes/MIA joint data\cite{prl}, the authors have
pointed out that the reconstruction of the shower longitudinal development
depends upon the assumed functional form of the features of the shower
development. How much do the characteristic parameters of the shower
longitudinal development such as the shower maximum and its location, shower
rise and shower decay constant depend on which function is employed?  In the
new technique of shower reconstruction developed for the HiRes monocular data
\cite{HR1}, the shower development function plays an even more fundamental
role in the shower geometry reconstruction. In effect, the shower geometry is
varied until the correct shower width predicted by the average shower function
is obtained.  The use of an appropriate shower development function is crucial
in the determination of shower energy because this is based on the integral of
this function. In practice, empirical functions based on data at lower
energies or based on theoretical electromagnetic cascade
calculation are used at the highest energies, e.g. above 10$^{17}$eV. However,
none of these has been experimentally tested at these energies in the
atmosphere.  Values of parameters used in trial functions
such as the $\lambda$ in the Gaisser-Hillas function\cite{gh} and $L_0$ 
(for a hadronic shower) in the
Greisen function \cite{Greisen} have never been measured in experiments at
such a high energy.  This is sufficient motivation to carefully test these well
known functions and determine those parameters directly from experimental
data. The HiRes/MIA joint experiment meets the necessary conditions for such a
high resolution measurement of the shower longitudinal development.

The HiRes prototype detector provides a degree by degree measurement of the
intensity of the fluorescence light produced by the shower electrons along the
shower axis. This light intensity, after corrections, is proportional to the
shower development profile.  The angular bin signals, however, suffer
fluctuations and are contaminated by the direct and scattered Cerenkov light
component produced by the same electrons. The source of the fluctuation is sky
noise and the statistical fluctuation due to finite sampling of the light.  A
method to minimize the bin by bin fluctuation is to take averages of the
angular bin signals over all the events. This will establish an almost perfect
``average shower longitudinal development profile" which can be used to
test the various trial functions.  In order to do so, however, all showers must
be ``aligned" and ``normalized" so that the showers can be compared with
each other in spite of variations of position of shower maximum and shower
energy.  To carry out the alignment and normalization properly requires that the shower
geometry must be well determined.  In addition, the Cerenkov components of the
signals must be subtracted.  Fluorescence light is proportional
to the shower size and the proportionality coefficient is geometry related. In
the HiRes/MIA joint experiment, the shower geometry is well determined with
the help of muon front timing information from MIA, 
the wide elevation angle range (up
to 70$^\circ$) of the HiRes prototype detector allows for a broad range of 
shower
development to be detected,  and  the short shower-detector
distance (about 3 km) minimizes uncertainties related to the atmospheric
attenuation and scattering.  We use the same data set selected
for the study of the composition of primary cosmic rays\cite{prl}. 
The criteria used to select this sample 
are well justified there based on a  full detector simulation \cite{apj}.

In the following sections, we will describe the experiment and data set 
briefly, explain how to minimize the contribution of Cerenkov light and how to
subtract the Cerenkov light component in each degree bin, propose a way to
normalize the individual showers, and present  the average profile 
and compare it with well-known functions.

\section{HiRes/MIA Joint Experiment and Data Set}

The HiRes prototype is situated at
$112^{\circ}\,\mbox{W}\,\mbox{longitude}$ and $40^{\circ}
\,\mbox{N}\,\mbox{latitude}$ and at a vertical atmospheric depth of $850$
g/cm$^2$.  It is\cite{hires} composed of 14 optical reflecting
telescopes. They image the extensive air shower(EAS) as it progresses through
the field of view of the telescope.  Nitrogen fluorescence light
(300--400\,nm) is emitted at an atmospheric depth $X$ in proportion to the
number of charged particles in the EAS at that depth. The electrons radiate
Cerenkov light in the same band simultaneously. Because of the very
foreward angular distribution of the Cerenkov light, only those events pointing
toward the detector will give rise to signals dominated by Cerenkov light. 
However, the
Cerenkov light can be scattered by molecular and aerosol particles into almost
all pixels along the shower track no matter what the shower geometry
is. Therefore, the signal recorded by the 
HiRes detector consists of the sum of 
fluorescence light, Cerenkov light and various kinds of noise.

The MIA detector is situated about 3.3 km from HiRes and 150 m lower in
elevation. It~\cite{casamia} consists of 2500 m$^2$ of active area distributed
in 16 patches of 64 scintillation counters and measures the EAS muon arrival
times with a precision of 4 ns and records all hits occurring within 4
$\mu\mbox{s}$ of the system trigger.  Combining such well determined timing from MIA
with the triggered HiRes pixel directions and HiRes timing information allows
precise reconstruction of the shower geometry. This is essential in the
determination of the atmospheric depth at which the shower passes the
triggered HiRes pixels and the optical path length  between the shower and the
detector.


HiRes/MIA coincident data were collected on clear
moonless nights between Aug. 23, 1993 and Aug. 24, 1996. The total coincident
exposure time was 2878 hours corresponding to a duty cycle of 10.9\%. 4034
coincident events were observed.  For events passing a set of coincidence
assurance cuts the shower trajectory, including arrival direction and core
location for each event was obtained in an iterative procedure using the
information from both HiRes and MIA \cite{Brian}.  The accuracy of the shower
axis determination depends on the number of observed muons, the HiRes angular
track length and the core distances from MIA and HiRes. 2491 events are
reconstructed via this procedure.  Monte Carlo(MC) studies \cite{apj} show that
the median shower direction error is $0.85^\circ$ with a median core location
error of 45\,m.

To insure data quality and maintain good resolution we require that for each
event the depth of shower maximum $X_{m}$ is visible within a minimum 
observed slant depth interval of 250
g/cm$^2$, that the track subtends at least 20$^{\circ}\,$, that the
accumulated gap between the fired pixels is less than 40\% of the
total gramage spanned by the shower, that the uncertainty in $X_{m}$, $\Delta
X_m$, is less than 50 g/cm$^{2}$, that the reduced $\chi^{2}$ for the
profile fit not exceed 10 and that the MIA to core distance
$R_{p_{MIA}}$ is less than 2000 m.  Additionally, we require 
the minimum viewing
angle of the fired tube, $\theta_h$, to be greater than 20$^{\circ}$ in
order to delete events which possess large direct Cerenkov light components. These
cuts, as summarized in TABLE 1, leave a sample of 488 events.  The energy
distribution peaks at 3$\times$10$^{17}$ and the 
average Cerenkov light fraction
is 15\%. There is no event in this sample in which 
direct Cerenkov light contributes more than 1\%
of the total signal.

We note that the shower development profile has been fit to the
Gaisser-Hillas(G-H) function in order to carry out some of the cuts.  However, the use of
this function as well as the fit parameters derived from it is for the purpose
of event selection only.  Hereafter, only the raw data for these selected
events is used in the analysis.
 
\section{Cerenkov Light Subtraction}
HiRes pixel signals are rearranged into a one dimensional series of one degree
bins along the shower track by considering the light acceptance of the mirrors
according to the geometric position of the shower. As mentioned above, these
HiRes bin signals include both fluorescence and Cerenkov light, although the
total fraction of direct Cerenkov light is kept 
low by the imposition of the cuts
described above. Bins which correspond to the deeper positions in the
atmosphere tend to have more scattered Cerenkov light because the Cerenkov beam
accumulates with increasing shower development and the molecular and aerosol
scattering gets stronger as the shower penetrates deeper into the
atmosphere. The raw signal versus depth, therefore, is not in itself a
representation of the shower longitudinal development.  We must
subtract the scattered Cerenkov light from the observed signal. The resultant 
corrected signal composed only of the fluorescence light contribution
is proportional to the shower size at the corresponding atmospheric
depth except for the natural and electronic noise component. 

    The Cerenkov light produced in an angular bin is calculated by multiplying
the average Cerenkov light yield of a single electron with the shower size at
the corresponding depth $X$. The average yield is taken over the energy of shower electrons
above the Cerenkov threshold, $E_t$, and over the emitting angle. The energy
spectrum of shower electrons, the angular distribution of Cerenkov light and
the pressure dependence of $E_t$ and yield are carefully considered. Details
can be found in the Ref. \cite{FEnim} and its references.

The contribution of direct Cerenkov light is minimized by the data cuts.
The contribution to the bin signal from scattered light from the
accumulated Cerenkov beam is estimated as follows: we assume that the Cerenkov
light beam in the angular bin is due to shower electrons in the previous
adjacent angular bin.  After
taking into account the attenuation of the Cerenkov light beam traveling
through the air between the bins, the contribution of scattered Cerenkov 
light from the beam into the
detector is estimated based on the Rayleigh scattering theory and Mie
scattering by aerosol.

 The observed signal in this angular bin can then be modified by subtracting
the scattered Cerenkov light. The corrected signal can be
converted into a shower size for the sake of computing the contribution to the
Cerenkov light beam in the next adjacent angular bin.  This recurrent
procedure requires knowledge of the Cerenkov light beam component in the first
angular bin in the field of the view of the detector. If the first signal bin 
were at the beginning of the shower development, the Cerenkov light component 
would be {\it zero}. The observed signal in this bin would be  pure fluorescence
which can be converted directly into a shower size.  
We assume that the first detected bin corresponds to an early enough stage 
in the shower development so that the zero-Cerenkov-light-component assumption
is still approximately  true.
This subtraction scheme systematically underestimate the Cerenkov
fraction in each subsequent bin because the first detected signal from the shower is not
at the very beginning of the shower cascade.  However, we will find  that
this systematic error is negligible and the proposed Cerenkov light
subtraction is sufficient for our purposes. 

\section{The Normalization of Air Shower Development Profile}
As mentioned above, all the showers must be aligned in atmospheric depth and
be normalized in shower size before one can take the average in bins over events.
Most of our events follow a transition curve, namely the shower size increases
rapidly as it develops in the atmosphere until reaches its maximum, $N_{m}$,
then it starts decreasing because shower electrons begin to lose more energy
by ionization than by radiation of high energy gammas.  The maximum size of a
shower is approximately proportional to the energy of the primary particle.
After subtracting the Cerenkov light components, the signals are converted
into shower sizes by taking the geometry and atmosphere related light
collecting efficiency and attenuation effect into account. The showers are
normalized with respect to energy by their maximum size, i.e.  $N(X)/N_{m}$,
denoted as $n(X)$. All the shower sizes are normalized to one as they reach 
their maxima.

    Air shower development fluctuates in atmospheric depth 
due to fluctuations in the hadronic multiparticle production 
(which depends on the nature of the primary nucleus) and electromagnetic
processes. As a consequence, the position of the shower maximum, $X_m$,
 varies from event to event. One can express
the shower longitudinal development as a function of the shower ``age",
defined as $s=\frac{3X}{X+2X_{m}}$, instead of depth, $X$. By using $s$,
the shower longitudinal development is universally described as a rising phase
from 0 (initial position of the shower) to 1 (shower maximum) and the decay
phase from 1 to 3 (infinite depth). 
The physical shower has an effective extent 
from 0 to 2, however.

    After the Cerenkov light subtraction, the observed signal profiles of
individual events span a range of atmospheric depth and show the
shower maximum through their convex-upwards shape. The best way to find the position of
shower maximum in the presence of statistical fluctuations is to carry out a
local fit with a rather general function.  We use a parabolic function to fit
the data in order to determine both the location and the size of the shower
maximum. The only constraint to this function is that it must be convex upwards. Here,
``local fitting'' means that the fitting procedure is only applied to those
data points near the maximum to minimize the bias in the location of the
maximum caused by using such a symmetric trial function. The method 
can be iterated to remove points far from the maximum. 
We find that the iteration affects the location of shower maximum by
less than 4\% however and we do not use it in what follows.

With shower maximum and size at maximum determined, the showers 
 are normalized in size
and aligned in depth. FIG 1 shows all 488 showers plotted with
normalized shower size, $n$, versus shower age, $s$.  In order to 
more easily see
the change of the density of the dots in the scatter plot, we let each entry
represents three measurements.
6069 entries in total out of 18230 bins signals are picked randomly and 
plotted. We note that there is a 20\% fluctuation in $n$ near the shower
maximum. The distribution of $n$ at shower maximum 
 is shown in FIG 2. The residuals in $n$,
defined as the ratio between the deviation from the expectation value and the
error associated with the individual bin signal, are well distributed as a
Gaussian centered at 0 with a width of 0.94. The main sources of this
fluctuation are sky noise, statistical fluctuation, the error in shower
geometry determination, the error in Cerenkov light subtraction and the error
in determining shower maximum. The first three, and particularly the first
one, contribute more than 80\% to the total fluctuation in the signal.  The error
from Cerenkov subtraction contributes the least.

    The data is binned into age intervals, instead of the original 
angular bin. We take an average of the normalized shower size within each
bin over all events to obtain an average shower development profile. 
The curve is well
determined between $s$=0.5 and 1.25.  We demonstrate the effect of the
Cerenkov light subtraction in FIG 3, by plotting the average shower
transition curves before and after the subtraction in the same plot.

\section{Tests of trial functions}
    After calculating the average normalized shower longitudinal development
profile, we fit  the shape of the profile with several well known trial
functions such as the Gaisser-Hillas\cite{gh}, Greisen\cite{Greisen} and Gaussian
forms.

    In order to describe the characteristics of the shower longitudinal
development, i.e. the asymmetric rising-falling shape, Gaisser and Hillas
introduced a function which depends on the essential characteristics an air
shower: the initial point, $X_0$, the shower maximum, $N_{m}$, the shower
maximum location, $X_{m}$, and the shower decay length, $\lambda$.  Except for
$N_{m}$ which is in number of shower electrons, all the parameters are in
units of atmosphere depth, $g/cm^2$. The function describes the change of the
shower size with the atmospheric depth as 
\begin{eqnarray} 
N(X) = N_{m} \left(
\frac{X-X_0}{X_{m}-X_0} \right)^\frac{X_{m}-X_0}{\lambda}
e^\frac{X_{m}-X}{\lambda}\,, 
\label{1} 
\end{eqnarray} 
where the atmospheric depth, $X$, is in $g/cm^2$.

Translating the depth $X$ into age $s$ and using the normalized shower size 
$n=N/N_m$, eqn. (1) becomes  
\begin{eqnarray} 
n(s) = \left(1
- \frac{1-s}{3-s}\frac{3T_m}{T_m-T_0}\right)^{T_m-T_0}
e^{3T_m\frac{1-s}{3-s}}\,, 
\label{2} 
\end{eqnarray} 
where $T_m=X_m/\lambda$
and $T_0=X_0/\lambda$ are the two remaining parameters. Parameter $T_0$ is
constrained to be less than $\frac{2s_{min}}{3-s_{min}}T_m$, where $s_{min}$ is
the lower limit of the data points (about 0.5 as shown in FIG. 3).  The
comparison of this function with the data is shown in FIG. 4.  
The two parameters are strongly correlated.  

The fit is poor at small and large $s$, particularly beyond $s=1.3$. 
While a possible reason for the deviation of those 3$\sim$4 points may be  
poor statistics, a detailed systematic error 
analysis is described in the next section.

     The second trial function is the Greisen function which was
suggested\cite{Greisen} to describe the development of a pure electromagnetic
shower. Depth is therefore expressed in the unit of radiation length,
$L_0=36.66g/cm^2$. The function is defined by a single parameter, 
$y=X_m/L_0$, and has a form
\begin{eqnarray} 
N(T) = \frac{0.31}{\sqrt{y}}e^{T(1-\frac{3}{2}{\it ln}s)}\,, 
\label{3} 
\end{eqnarray}
where $T=X/L_0$ refers to the atmospheric depth in radiation lengths in
 the air. Since the age $s$ appears
explicitly in the formula, it is straightforward to rewrite it as a
function of either $s$ or $T$. It is apparent after conversion to a 
function of $s$
that the radiation length no longer appears explicitly in the formula.
This feature of the Greisen function make it 
potentially useful for the description 
of hadronic shower with a single parameter $y$. However, the meaning of $L_0$
need to be changed. In the case of electromagnetic cascade, Greisen shows 
that $y={\it ln}(E_0/E_c)$ fits a elongation law with a rate about 60g/cm$^2$
where the $E_c$ refers to the electron critical energy. We will see how the 
rate is changed to fit the case of hadronic shower below. The fit of the 
Greisen function to our measured average shower is shown in the dashed
line in FIG. 4.  

     In the Fly's Eye data analysis\cite{FEcomp},  
the poorer detector resolution dominated over the shape of the 
shower longitudinal development and a symmetric Gaussian in $X$ 
could be used to represent the shower longitudinal development. 
Using $\xi=\sigma/X_m$ as the single parameter, the Gaussian can be
normalized by following the rule set in previous sections, i.e.
\begin{eqnarray} n(s) = exp\left\{ -\frac{1}{2\xi^2}\left(
 \frac{1-s}{1-s/3} \right)^2 \right\}\,.
\label{4}
\end{eqnarray}
The result  is also plotted in FIG 4.  It is  a poor  fit to the data
with a $\chi^2=26$ per degree of freedom. Thanks to 
the significantly improved resolution, this experiment 
provides the first clear observation of an asymmetrical  air shower 
development profile in $X$ at 10$^{17} \sim$ 10$^{18}$eV.  

However, based on the result of this experiment, we observe that the average
longitudinal development profile of air showers appears quite symmetrical as a
function of shower age, $s$. A simple symmetrical Gaussian shaped
function of age can be used 
to describe shower profiles.  Since the function is centered
at 1, it is governed by only one parameter, i.e. the width of the function,
denoted as $\sigma$. This fit is  shown in FIG. 4 also. Hereafter, 
we call this function the ``new Gaussian function" 
to distinguish it from the one that is symmetric in the depth variable.

We plot the fit results in the form of deviations in FIG.
5. Quantitatively, the $\chi^2$ per degree of freedom for 
Gaisser-Hillas, Greisen and the ``new 
Gaussian function" are 1.93, 1.87 and 1.79 respectively. 
They all have essentially the same goodness of fit. More data beyond the region
currently covered is required to further differentiate between them. 
From a practical standpoint however, the
Gaisser-Hillas function contains too many parameters, especially $X_0$ 
which is not directly measured in experiments and is consequently poorly defined. 
These parameters are strongly correlated. The Greisen function 
does not have this problem because it fits the data
with a single parameter.
Both functions need a scale to measure the shower development, however. 
The scale is given by $\lambda$ and $L_0$, respectively. The scale must be 
associated with the type of primary particle. This actually reduce the 
freedom of fitting in the reconstruction of
real experimental events. One has to set the scale as a parameter in a
complete fitting procedure. The ``new function" proposed in this work is more
practically useful in real shower reconstruction. As a function of depth, it
has the form  
\begin{eqnarray} 
n(X) = exp\left\{
-\frac{2}{\sigma^2}\left(\frac{X-X_m}{X+2X_m}\right)^2\right\}\,.
\label{5}
\end{eqnarray} 
This function uses a single parameter, $X_m$, to locate a shower
in the atmosphere, and a single parameter, $\sigma$, to indicate the shower 
width and it is {\it scale-free}. The last feature is very useful 
since the composition is a-priori unknown.

\section{Systematic error analysis} 
In order to investigate the reasons 
for deviation from the test functions at both early and late stages of 
shower development, we apply the above analysis to Monte
Carlo simulated events. The generator used here is the full simulation code
of the HiRes/MIA detector developed in the 
composition study\cite{prl}. It is driven by a shower generator based on 
the CORSIKA package. It produces the shower size using the Gaisser-Hillas
function with  full fluctuations in $X_m$, $N_m$, $X_0$
and $\lambda$. The fluctuation in muon density and muon arrival time on the 
ground is generated according to parameterizations of 
CORSIKA simulated events. All details of the detector are carefully 
considered including the sky noise, atmospheric attenuation,
``ray-tracing", phototube response, electronics and triggering. Generated
data are passed through the complete calibration and reconstruction 
procedure just like the real data. The details of the generator and the 
comparison with the data can be found in the Ref. \cite{apj}. FIG. 6 shows that the
overall quality of the fit is good. This means that the Cerenkov
subtracting scheme, shower maximum finding and normalization procedure
proposed in this paper work well.  The shower longitudinal development
function is  reconstructed almost exactly except for the few points at the
beginning and the end of the shower as is the case for the real data. The
deviation turns out to be associated with the correction to tube signals which
lie far from the shower-detector-plane.  Details of this correction can
be found in Ref. \cite{Brian}.  Briefly, a tube can be triggered even if it is
not located exactly on the shower-detector plane since the shower has a finite
lateral width, the mirror image on the focal plane has a finite spot size and
the broadening effect due to the atmospheric scattering. The
amount of light reaching the detector can be calculated by 
correcting the amount of measured light with a response function in
 the off-plane-angle. This function is determined by performing a
careful ``ray-tracing'' procedure which folds all of the effects in. It turns
out to be a Gaussian with a typical width of a few tenths of a degree depending
on the shower distance and the shower age. On the other hand, we know that
there is an uncertainty in the determination of the shower-detector plane 
direction of about  0.7 degrees\cite{apj}. 
This uncertainty causes a systematic overcorrection in the signals for 
off plane tubes. This becomes
worse for a tube located far from the center of the phototube cluster.  This
effect can be verified by simply substituting
 the ideal MC input shower-detector
plane for calculating the correction and comparing with the correction using
the fitted shower-detector plane. This is shown in FIG. 6. The effect on
the beginning and end of the average profile is clear. 

One way to handle this systematic error would be to correct it based on this
MC simulation result. Another approach is to simply drop those
affected points in the fitting procedures. We prefer the later for several
reasons. First, it avoids introducing any modeling dependence. Secondly, the
points being dropped at age $>1.25$, correspond to positions deeper than
1.43$X_m$, or about 860$g/cm^2$ in the atmosphere, 
where the uncertainty in atmospheric attenuation  is 
greatest. As a part of the systematic uncertainty study, we varied the
atmospheric parameters\cite{prl} by one standard deviation from their most
probable values in the shower reconstruction. We observed that there is no
effect for most of the average profile and that this variation affected 
only the last few deepest points.    This effect is smaller than that
produced by the shower-detector plane error.  Finally, in practice,
dropping  those points  does not change our result because of
the relatively large statistical errors associated with those points as 
can be seen in  comparing
the FIG. 4 with FIG. 5, where events associated with the last three points 
in FIG. 4 are dropped. 

We now discuss the systematic error caused by setting the Cerenkov light to be
zero in the first angular bin in the field of view of the detector. 
Based on the previous results, we
known that the Gaisser-Hillas function works well in
describing the shower longitudinal development.  We can use the Gaisser-Hillas
function in a least $\chi$-square fitting procedure event by event. Since in
this case we know the shape of the shower, the Cerenkov light component in all angular
bins can be calculated.  The fraction of Cerenkov
light in each bin is compared to our original method in FIG. 7. It is clear
that most of the Cerenkov light is subtracted with our bin signal based
Cerenkov light subtracting scheme. Only 7.7\% of the Cerenkov light is
systematically underestimated. This is a minor effect in calculating the 
average transition
curves as shown in the FIG. 3. The diamonds in the figure are obtained by
carrying out the same averaging and normalization procedure as before 
but with the Cerenkov light fraction in each bin replaced by that 
determined using the Gaisser-Hillas function.  The points are
visibly lower which makes the profile appear slightly wider. However, the
widening is not large enough to change our conclusion on the shape of the
longitudinal development function. For instance, the values of fitted 
parameter $T_0$ and $T_m$ are -2.57 and 8.6, respectively, which are consistent 
with the result listed below. 

The other systematic issue is the reliability of our maximum searching by
using local fitting with a parabolic function.  
The uncertainty associated with this method is
demonstrated by comparing the maxima with those obtained from the
reconstruction procedure using the Gaisser-Hillas profile.  FIG. 8 a)
shows the comparison between the $X_{m}$'s and FIG. 8 b) shows the difference between
the $N_{m}$'s.  The other way to illustrate the uncertainty is to compare the
resolution functions in $X_m$ and $N_m$ from simulation for both methods. It turns out the
resolution in $X_m$ for local parabolic fitting is 16\% worse than the standard
Gaisser-Hillas approach, and the resolution in $N_m$ is comparable in both cases.

\section{Results and conclusions}
     In order to extract the best values for the
parameters, $T_m$, $T_0$, $y$ and $\sigma$, in the Gaisser-Hillas, Greisen
and the ``new Gaussian  function", a well reconstructed data set is used.
The parameters are determined to be
\begin{eqnarray}
T_m & = & 9.7 \pm 2.0 \,,\\
T_0 & = & -3.2 \pm 2.9 \,,\\
y & = & 11.11 \pm 0.34 \,,\\
\sigma & = & 0.272 \pm 0.002\,.
\end{eqnarray}
A strong 
correlation between $T_m$ and $T_0$ is found for the
Gaisser-Hillas function. This correlation can be expressed as 
$T_0\sim -0.10T_m^2+0.74T_m-0.10$ and is shown in FIG. 9. 

These results imply: 1) the strong correlation between
parameters in the Gaisser-Hillas function means that the data can be fit
with a simpler form. Alternatively one can fix one of those parameters and apply the
correlation between them in the shower reconstruction; 2) The parameter
$\lambda$ in the Gaisser-Hillas function can be estimated to be $67 \pm 11
g/cm^2$ by using the average $X_{max}=651.9 \pm 3.5 g/cm^2$;
3) the parameter
$T_0$, or $X_0$, is likely to be a negative number in any individual shower 
fit. It is thus difficult to
interpret as the first interaction point of a shower. 
4) the parameter $y$, defined as $X_m/L_0$ in the Greisen
function, is close to $T_m$, defined as $X_m/\lambda$. This 
implies that the best value of the parameter $L_0$, $58.7 \pm 2.0
g/cm^2$, is no longer the radiation length. This is to be expected since
it is used to describe a hadronic shower; 5) a symmetrical function 
in age is a suitable representation of the shower longitudinal profile. 
This function is depth-scale invariant.

In summary, the shape of the extensive air shower longitudinal development is
investigated in the energy range from 10$^{17}$ to 10$^{18}$eV with the
HiRes/MIA hybrid experiment.  The profile is quite symmetrical as a function of
the age of the shower.  The ``new Gaussian function'' 
(\ref{5}), the Greisen function
(\ref{3}) and the Gaisser-Hillas function (\ref{2}) describe the shower shape
almost equally well, with the $\chi^2$'s of the test 1.79, 1.87 and 1.93  
respectively.  This is the first direct measurement of the shower  average 
longitudinal development
profile at these energies and covering such a wide range of shower age.
  
\section{Acknowledgements}
We acknowledge the assistance of the command and staff of Dugway Proving 
Ground. We are greatly grateful for the discussion with James W. Cronin 
and Paul Sommers. This work is supported by the National 
Science Foundation under contract No. PHY-93-21949, PHY-93-22298 and 
the U.S. Department of Energy. 


\begin{table}
\begin{tabular}{|c|c|}
\hline
Variables & cuts \\ 
\hline
\hline
$X_m$ &  $X_l < X_m < X_h$\\
\hline
Track Span & $X_h - X_l > 250$ (g/cm$^2$) \\
\hline
Track Angular Length & $>20^\circ$ \\
\hline
Total Gap in Gramage & $<0.4(X_h - X_l)$ \\
\hline
$\Delta X_m$ & $<50$ (g/cm$^2$)\\
\hline
$\chi^2$ & $<$10/DOF\\
\hline
$\theta_h$ & $>20^\circ$\\
\hline
$R_{p_{MIA}}$ & $<$2 km \\
\hline
\end{tabular}
\caption{Event selecting criteria. $X_h$($X_l$) refers to the depth corresponding to
the highest(lowest) bin in the field of view of the detector.}
\end{table}

\begin{figure}[b]
\label{fig1}
\epsfig{file=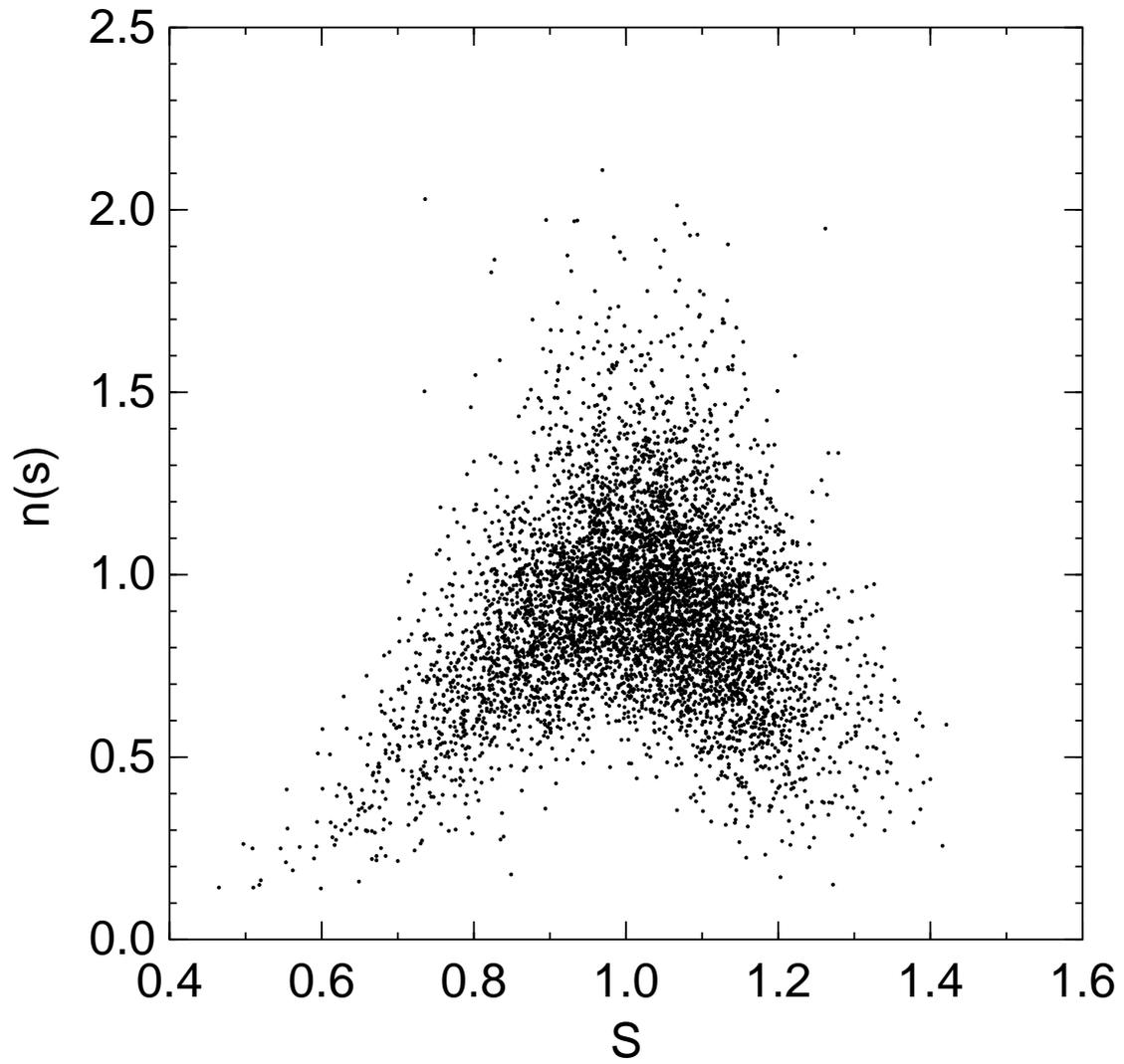} 
\caption{Aligned shower longitudinal development. Only one-third of the 
data is shown for clarity. Each point represents the
fluorescence light signal in an one degree bin in an event. 
The events are normalized to 1 at individual shower 
maximum via a local parabolic fit. }
\end{figure}
\begin{figure}[t]
\label{fig2}
\epsfig{file=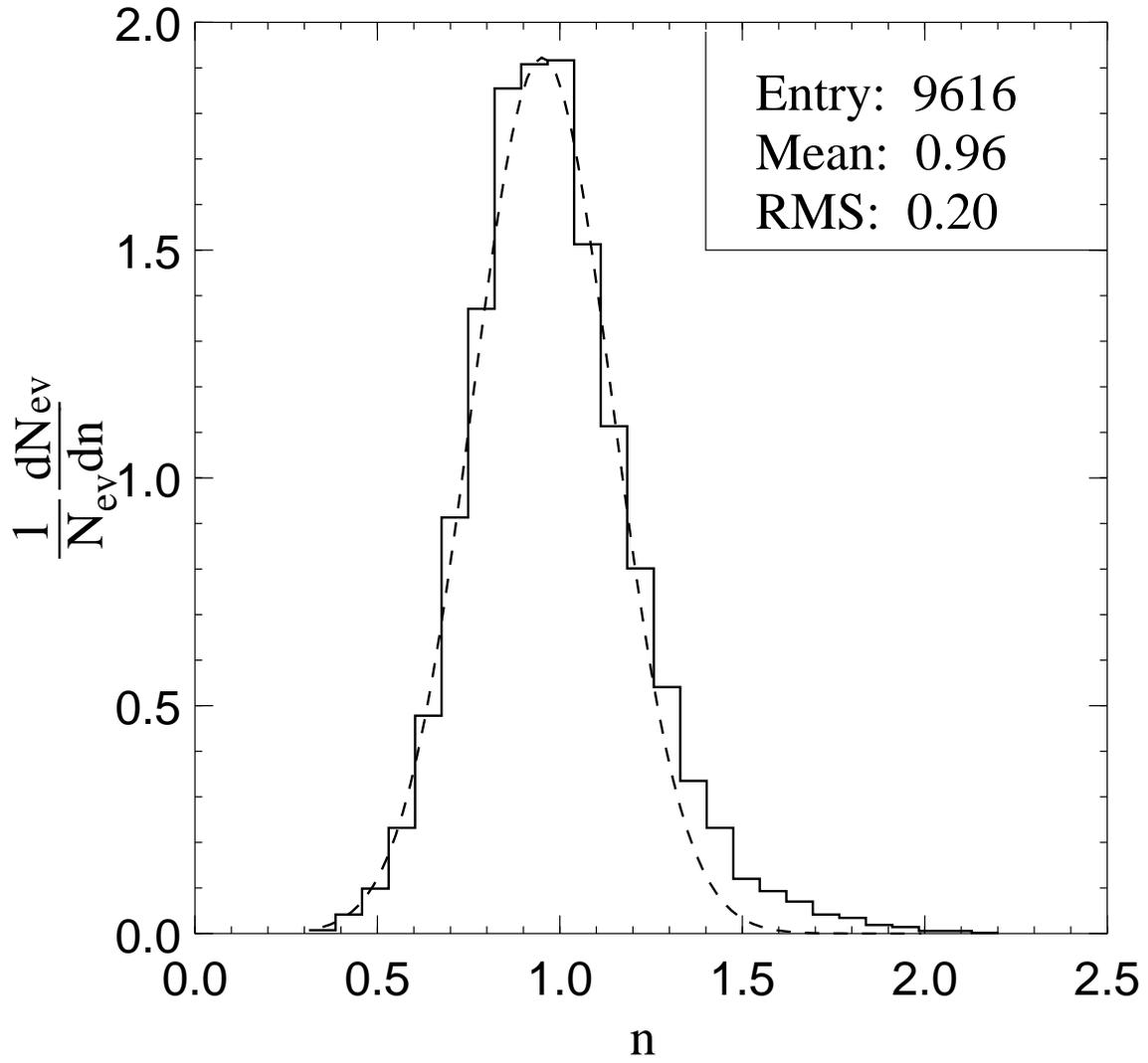}
\caption{The fluctuation of normalized 
fluorescence light signals around shower 
maximum, $0.9<s<1.1$. The dashed line represents a Gaussian fit, 
see the text for the details.}
\end{figure}
\begin{figure}[t]
\label{fig3}
\epsfig{file=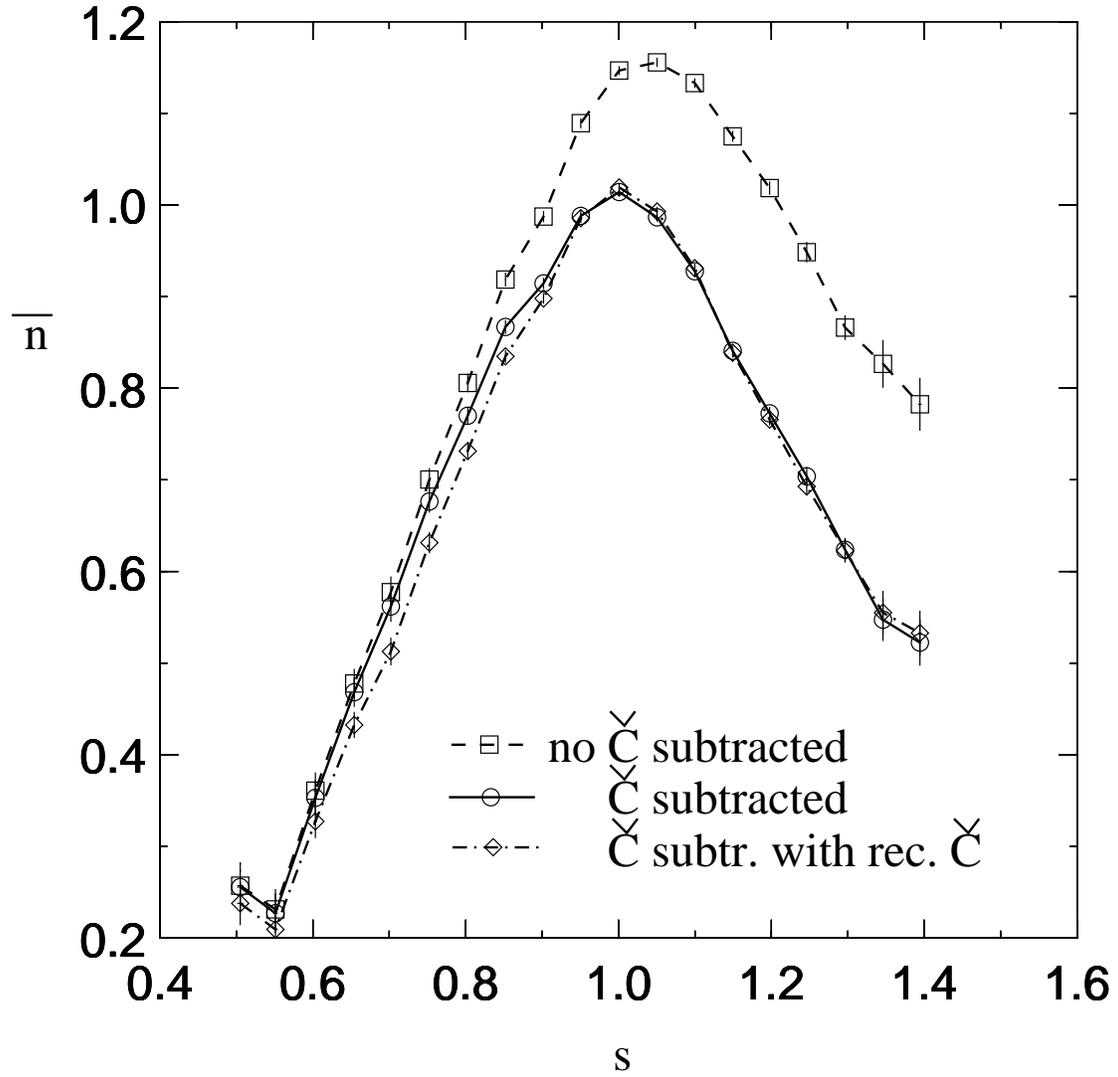}
\caption{The average shower transition curve. Solid line plus circles 
refers to the bin signal based recurrent Cerenkov-light-subtracting method, 
the dashed line plus squares is the raw signal including  
the Cerenkov light.
The dash-dotted line plus diamonds corresponds to a different 
Cerenkov light subtraction method (see the discussion
on systematic error in Section V). Lines are drown to guide the eye.}
\end{figure}
\begin{figure}[t]
\label{fig4}
\epsfig{file=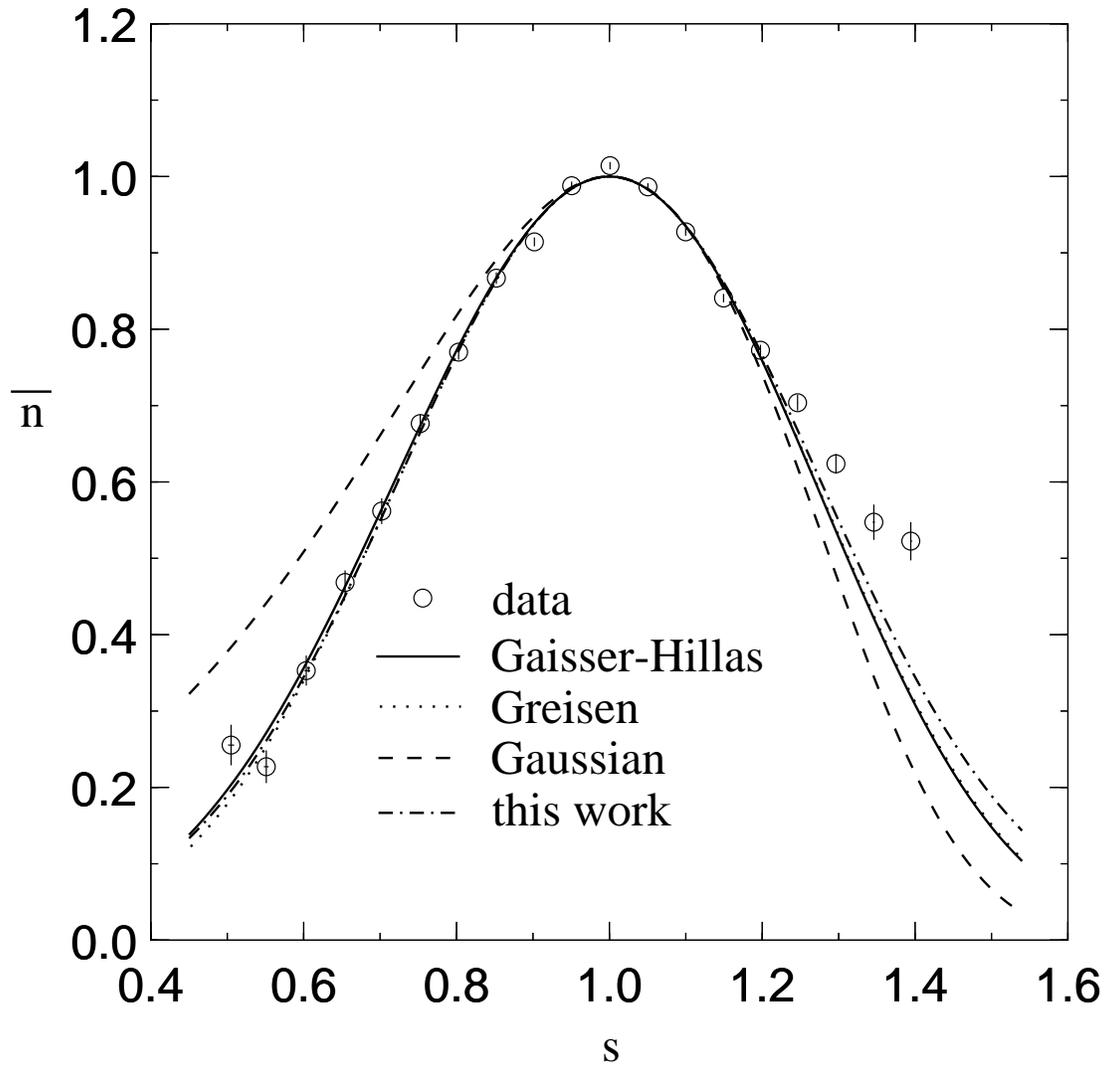}
\caption{Comparison between the data and test functions. Circles refer to  
data. The solid line
refers to the G-H function, the dotted line to the Greisen function,
the dashed line to the Gaussian function and the dash-dotted line to 
the newly proposed symmetrical Gaussian function of the shower age. }
\end{figure}
\begin{figure}[t]
\label{dev}
\epsfig{file=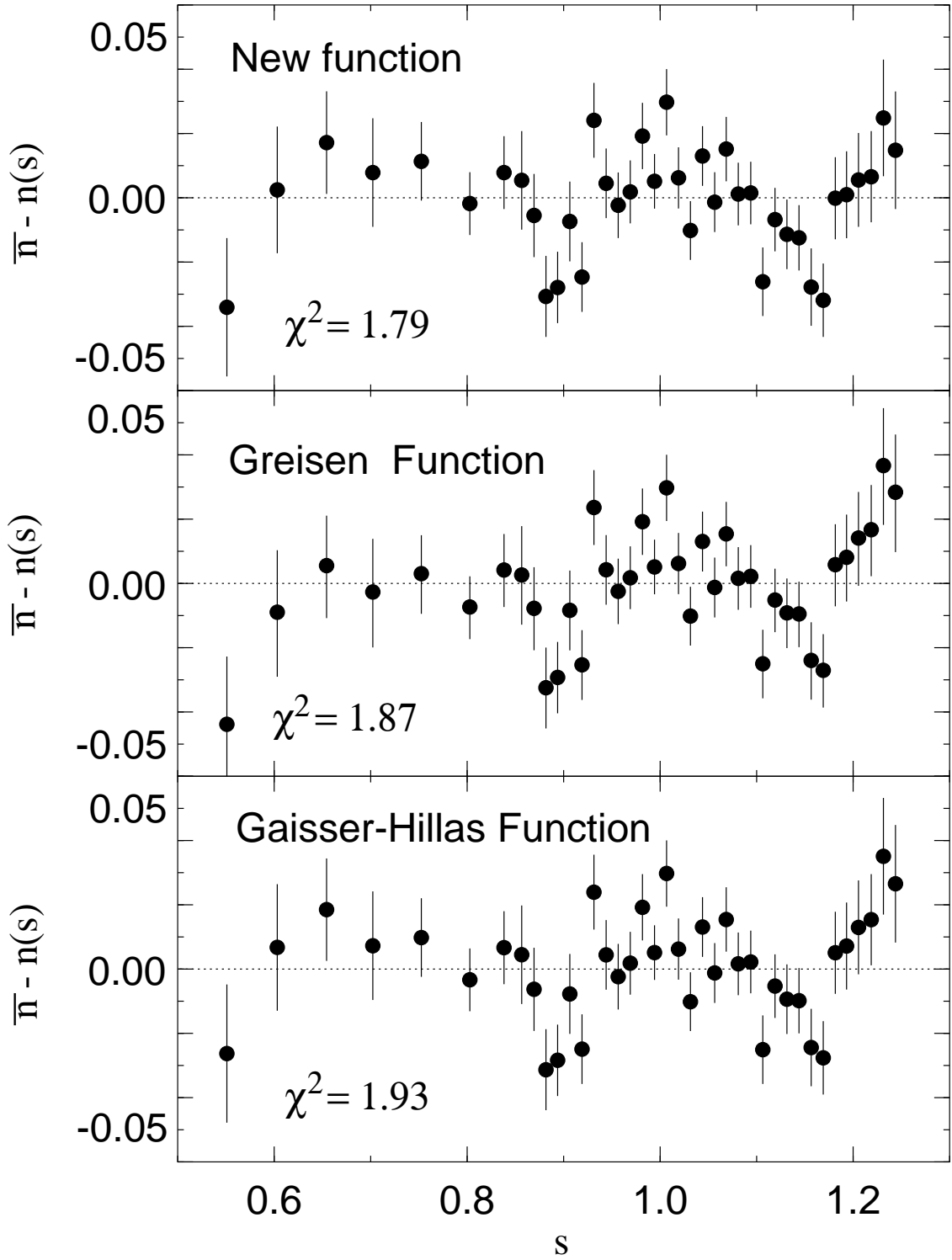}
\caption{Deviation of the data from the fits. The $\chi^2$ is per
 degree of freedom. } 
\end{figure}
\begin{figure}[t]
\label{mc}
\epsfig{file=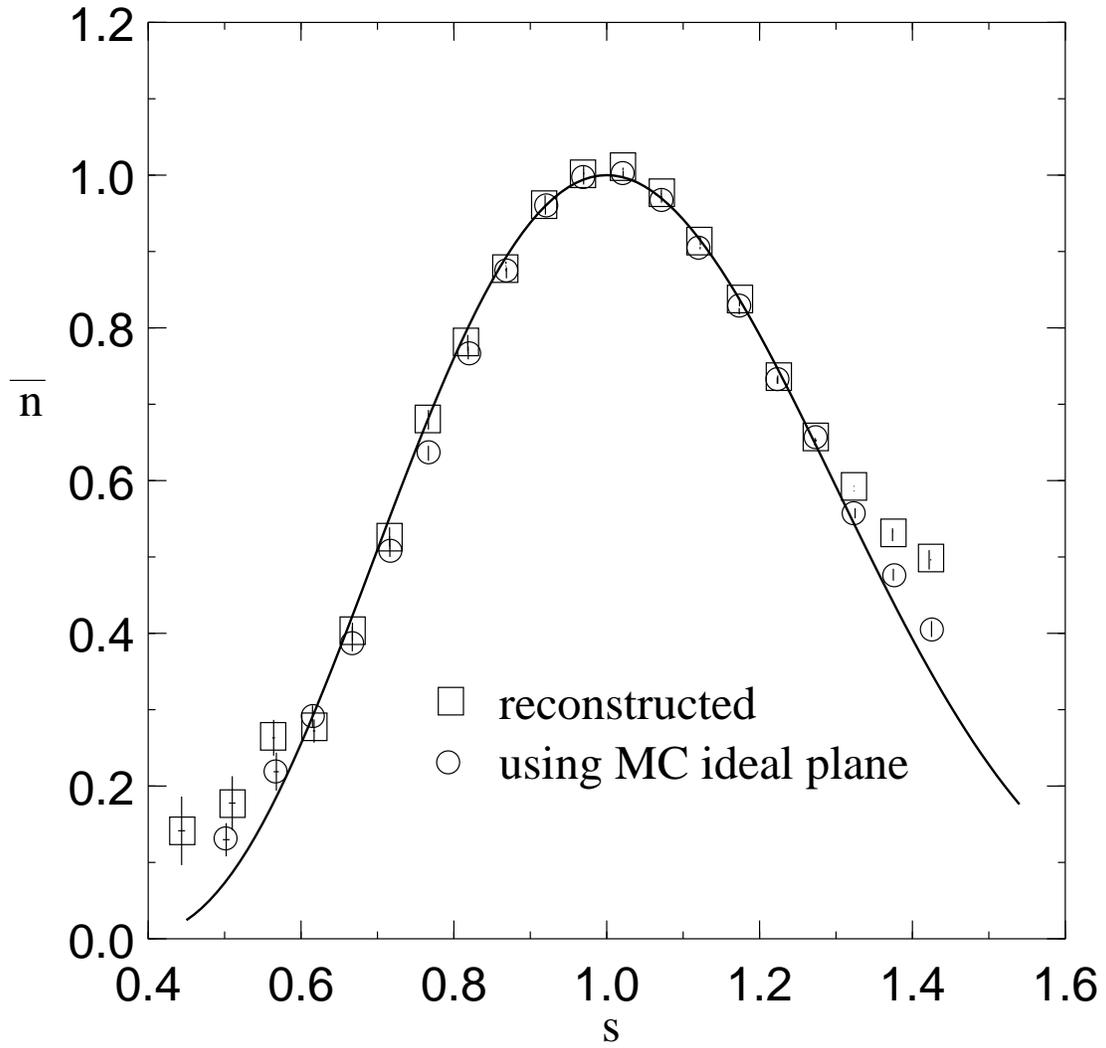}
\caption{The average longitudinal profile of MC generated showers. Open
squares represent the results from the standard shower reconstruction;
the open circles represent the same analysis but with the shower-detector
plane direction used correcting the 
tube signal replaced with the ideal input parameters.} 
\end{figure}
\begin{figure}[t]
\label{fig5}
\epsfig{file=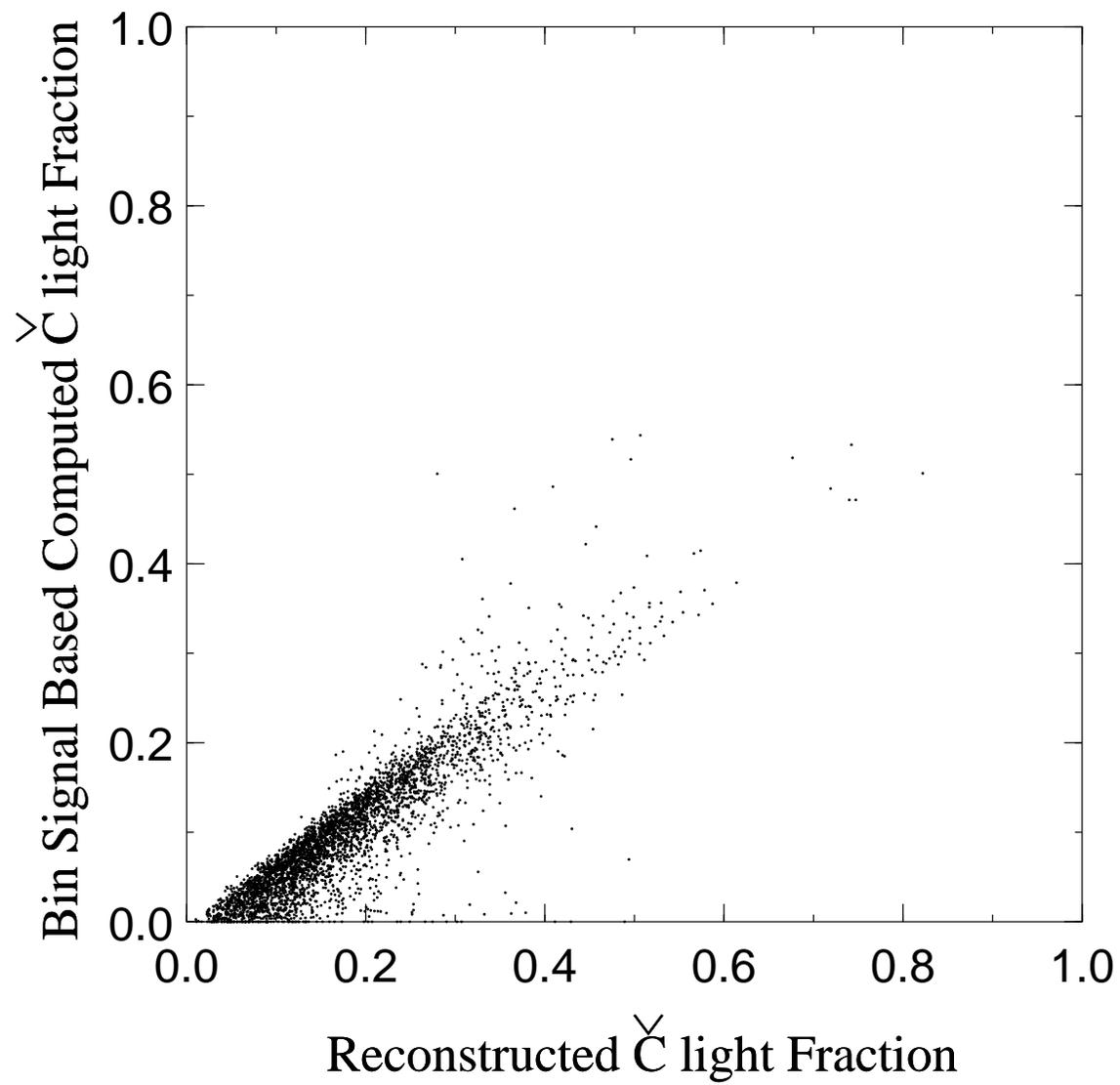}
\caption{Comparison of Cerenkov light fraction in each one 
degree bin. The x-axis is the 
reconstructed Cerenkov light fraction using G-H function and the y-axis
 is the fraction from the bin by bin recurrent Cerenkov-light-subtracting scheme.} 
\end{figure}
\begin{figure}[t]
\label{fig6}
\epsfig{file=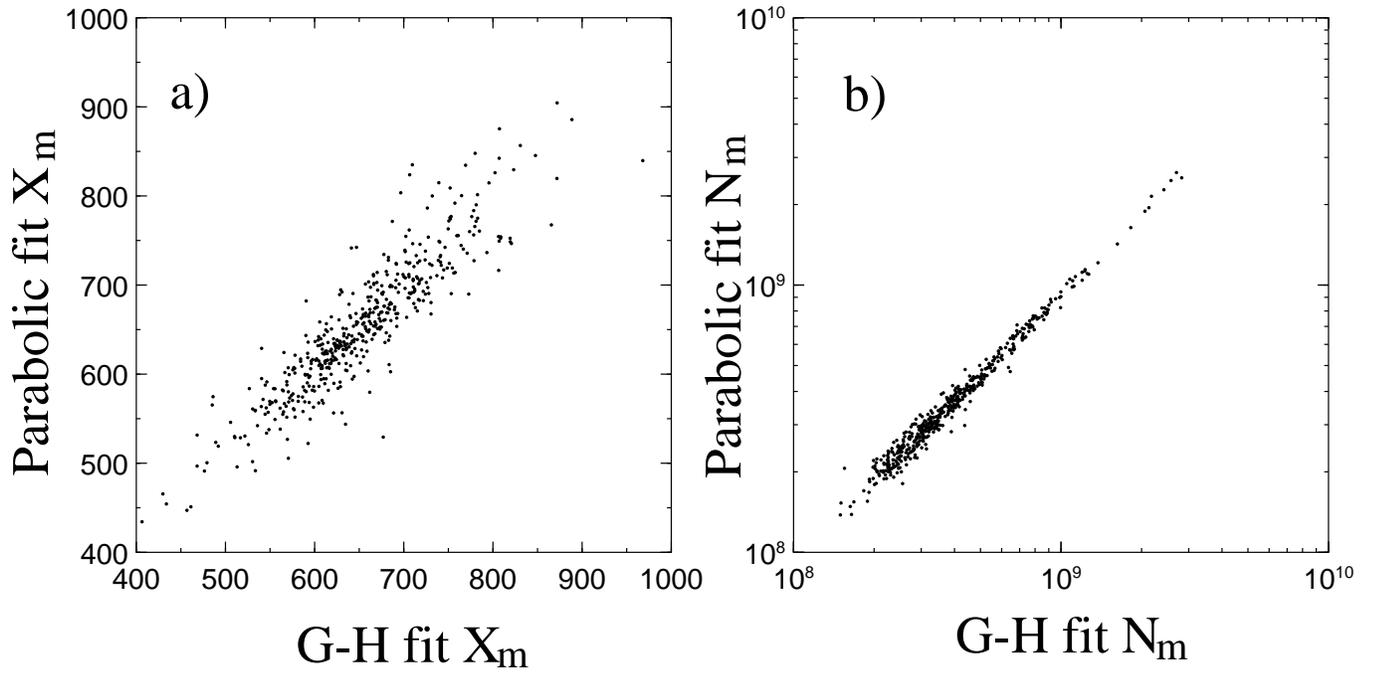}
\caption{Comparison of shower maxima for  
reconstruction with the G-H function and with local parabolic fitting.
a) for $X_m$'s and b) for $N_m$'s.}
\end{figure}
\begin{figure}[t]
\label{fig7}
\epsfig{file=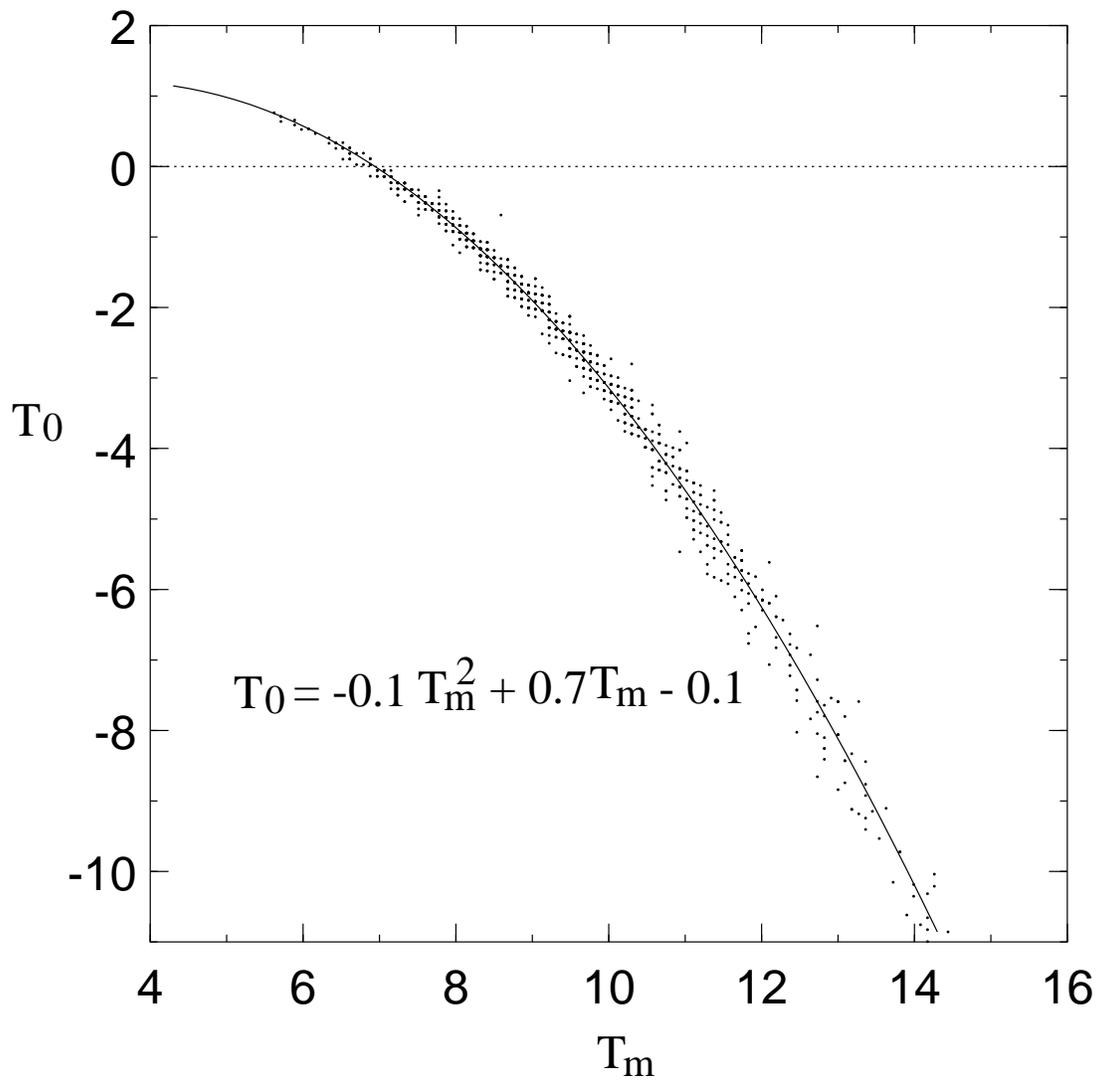}
\caption{The correlation between the parameter T$_m$ and T$_0$ in G-H function (\ref{2})}
\end{figure}
\end{document}